\documentclass[twocolumn,aps,prb,amsmath,amssymb, floatfix]{revtex4}
\usepackage[dvips]{graphicx}
\usepackage{dcolumn}
\usepackage{bm}
\usepackage{color}
\usepackage{soul}
\begin{document}
\pagenumbering{arabic}
\title{Switching energy-delay of all spin logic devices}
\author{Behtash Behin-Aein*, Angik Sarkar*, Srikant Srinivasan* and Supriyo Datta\\
\small* The order is alphabetical. Authors contributed equally.}
\normalsize \affiliation{School of Electrical and Computer Engineering
and NSF Network for Computational Nanotechnology (NCN)
Purdue University, West Lafayette, IN 47907}
\begin{abstract}
The need to find low power alternatives to digital electronic circuits has led to increasing interest in alternative switching schemes like the magnetic quantum cellular automata(MQCA) that store information in nanomagnets which communicate through their magnetic fields. A recent proposal called all spin logic (ASL) proposes to communicate between nanomagnets using spin currents which are spatially localized and can be conveniently routed. The objective of this paper is to present a model for ASL devices that is based on established physics and is benchmarked against available experimental data and to use it to investigate switching energy-delay of ASL devices.
\end{abstract}
\maketitle
\indent  
Digital electronic circuits store information in the form of capacitor charges that are manipulated using transistor-based switches. Switches of this type currently operate with a supply voltage of one volt involving $\approx  10^{4}-10^{5}$ electrons, requiring
$1-10$ femto-Joules (fJs), dissipating 1-10 $\mu W$ per switch if operating at 1 GHz. This dissipation per switch is believed to be the single most important impediment to continued miniaturization and there is a serious attempt to ``reinvent the transistor''\cite{Theis} so as to operate at lower voltages.\\
\indent  
A more radical approach is to replace the entire charge-based architecture with an architecture based on some other state variable such as spin\cite{Nikonov}. For example,  MQCA\cite{Imre} uses nanomagnets to represent digital information (0 and 1). Recently an all spin logic (ASL) device\cite{BehinAein} has been proposed whereby information is similarly stored in nanomagnets but is communicated via spin currents that are spatially localized and can be conveniently routed within a spin-coherence length which can be 100's of nanometers\cite{vanWeesOtani} to microns\cite{AppelbaumTombros}.\\
\begin{figure}[t]
 \centering
  \includegraphics[width=6cm, height=7.2cm]{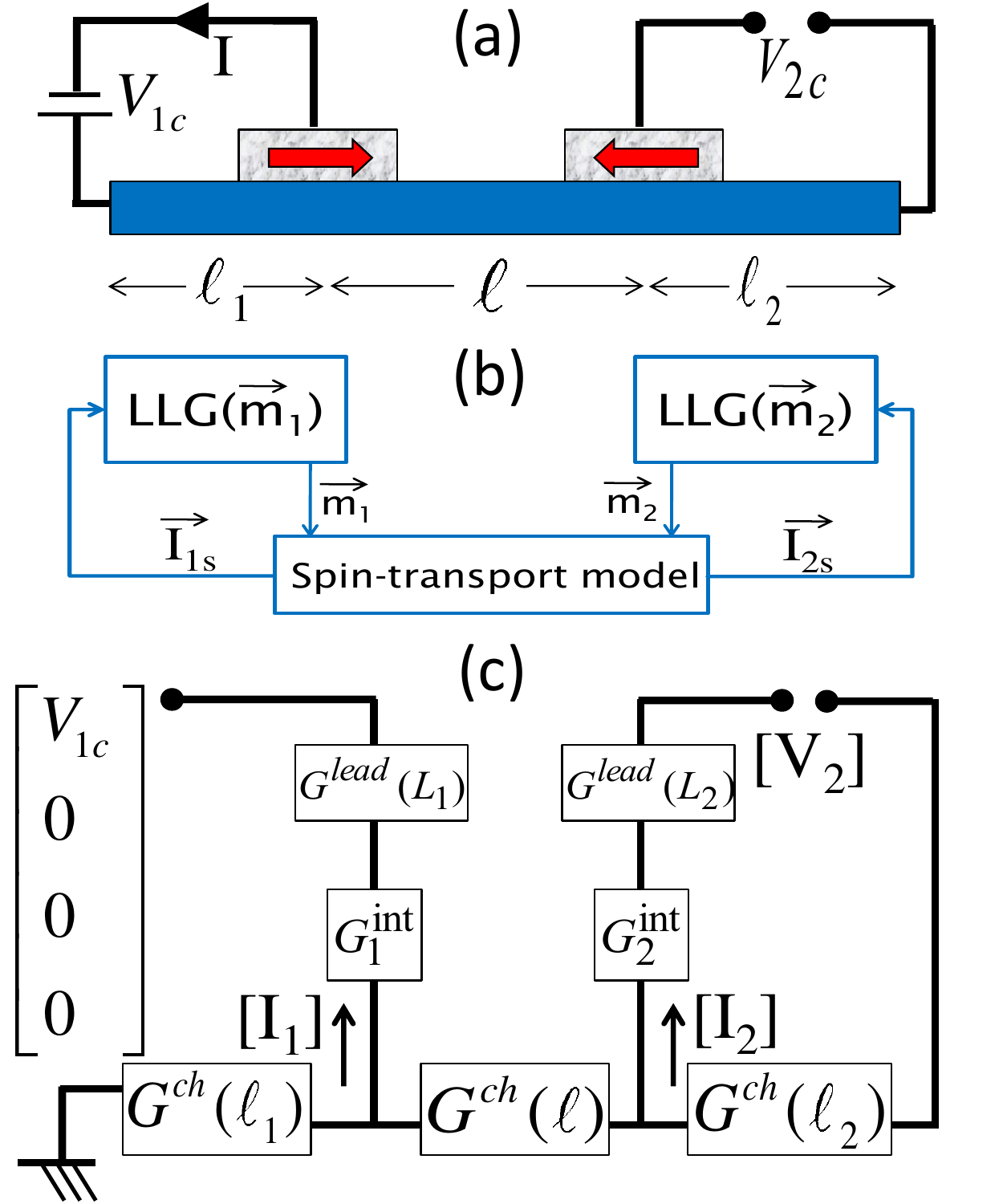}\\
  \caption{(a) An ASL device consisted of input and output magnets. (b) Illustrates the self-consistent model. (c) Shows the conductance matrices describing the spin-transport.}
  \label{fig1}
\end{figure}
\indent  
It has been argued that ASL devices could potentially lead to ultralow power switches since a stable nanomagnet with an activation barrier of $40$ kT could be switched with less than an attoJoule (aJ)\cite{BehinAein}. Experimentally, however, nanomagnet memory devices typically require tens of fJs to switch at speeds that are a factor of 100 to 1000 lower, raising questions about the potential of ASL devices to provide a low-power alternative to today's transistors. This is because most of the dissipation in switching magnets is associated not with the dynamics of magnets but with the spin transport process and we need a suitable model that incorporates both to make reliable predictions. This paper presents such a model that is based on established physics and is benchmarked against the recent experimental result of Yang et al.\cite{Yang}.\\
\indent 
In general, the switching energy and energy-delay can be written as:
\begin{equation}
\label{SW-ED}
E_{sw} = V\cdot Q_{tot} \hspace{0.25cm},\hspace{0.25cm} E_{sw} t_{sw} = \frac{V}{I}Q^2_{tot}
\end{equation}
$V$ and $I$ are the charge voltage and current respectively and $t_{sw}$ is the switching delay. $Q_{tot} = It_{sw}$ is the total charge involved in a switching event. Equation 1 permits a simple comparison with charge-based devices like today's transistors where $Q_{tot}$ is the amount of charge being switched. ASL devices permit low voltage operation especially if metallic channels are used. For example, the experiment in Ref.[7] uses a switching voltage of $\approx30$ mV and it is of the same order or less for GMR devices, far lower than today's transistors, and one objective of this paper is to use our quantitative model to provide insight into the factors that determine $Q_{tot}$.\\
\indent  
A generic ASL device is shown in Fig.\ref{fig1}a, with charge current going through an input magnet ($\vec{m}_1$) and an accompanying spin-current resulting in spin-torque\cite{Slonczewski} which if large enough could flip the output magnet ($\vec{m}_2$). Analyzing such a device involves coupling (Fig.\ref{fig1}b) a model for magnetization dynamics described using the Landau-Lifshitz-Gilbert (LLG) equation with a spin transport model (Fig.\ref{fig1}c). For the latter, we adopt what we could call a ``spin-circuit'' approach by combining (a) the well-established spin-diffusion model developed by Johnson-Silsbee\cite{Johnson} and Valet-Fert\cite{Valet} that are now widely used\cite{vanWeesOtani} for spin transport in long channels, with (b) a conductance model for the channel-magnet interface pioneered by Brataas et. al\cite{Brataas}, whereby a 4-component voltage drop is related to a 4-component current by a $\left[4\times4\right]$ interface conductance matrix:
$\left[I_c,I_z,I_x,I_y\right]^T = [G]_{4\times4}\left[\Delta V_c,\Delta V_z,\Delta V_x,\Delta V_y\right]^T$.
The four-components represent the charge `c' and the three spin components $z$, $x$ and $y$. Figure \ref{fig2} is plotted using (a) and (b) along with LLG (see also supplementary section S.1B), which shows the output voltage per unit current and agrees well with experimental data\cite{Yang}. To obtain a close match, polarization of the magnets was adjusted to a value of 0.5, which is in a reasonable range\cite{vanWeesOtani}. (S.1B,C provide the parameters used).\\
\indent  
The spin-transport model in Fig.\ref{fig1}c illustrates three basic ingredients: (I) the lead-nanomagnet-channel interface: $G^{int}$, (II) nonmagnetic-lead conductance: $G^{lead}$ and (III) the channel: $G^{ch}$. For an interface between a non-magnetic channel and a Ferromagnet pointing along $z$, \textbf{(1)} can be modeled as a conductance  matrix whose components can be written in terms of the scattering matrix between the plane `C' inside the channel and the plane `L' inside the lead. It has been shown\cite{Brataas} that:
\begin{equation}
\label{Gint}
G^{\text{int}} =
\left[
  \begin{array}{cccc}
     g  &  g P& 0  & 0  \\
     g P&  g  & 0  & 0 \\
     0  &  0  & \Gamma+\Gamma^* & i\left(\Gamma - \Gamma^*\right) \\
     0  &  0  & -i\left(\Gamma - \Gamma^*\right) & \Gamma+\Gamma^* \\
  \end{array}
\right]
\end{equation}
where $g = 2 - r_u r_u^* - r_d r_d^*$, $gP = r_d r_d^* - r_u r_u^*$, $\Gamma = 1 - r_u r_d^*$ and P
is the polarization. $r_u,r_d$ being the reflection coefficients for up and down spins respectively as seen from
the plane `C' inside the channel.  This is for a single conduction mode. All modes have to be added together to represent the interface area and materials used (S.1C).\\
\indent 
For (II) i.e. $G^{lead}$ we construct the full conductance matrix
for the contact by placing the interface conductance in series with a $\Pi$-conductance network whose series (se) and shunt (sh) components are given by ($\rho$: resistivity, $\ell$: length , $A$: cross-sectional area, $\lambda$: spin-flip length):
\begin{equation}
\nonumber
G^{\text{sh}}=
\left[
  \begin{array}{cccc}
     0  &  0 &  0  & 0 \\
     0  &  g_{sh}  & 0  & 0 \\
     0  &  0  & g_{sh}  & 0 \\
     0  &  0  & 0  & g_{sh} \\
  \end{array}
\right],
G^{\text{se}} =
\left[
  \begin{array}{cccc}
     \frac{A}{\rho\ell}  &  0 &  0  & 0 \\
     0  &  g_{se}  & 0  & 0 \\
     0  &  0  & g_{se}  & 0 \\
     0  &  0  & 0  & g_{se} \\
  \end{array}
\right]
\end{equation}
where $g_{se}\equiv \left(A/\rho\lambda\right)\csc h\left(\ell/\lambda\right)$
and $g_{sh}\equiv \left(A/\rho\lambda\right)\tanh\left(\ell/2\lambda\right)$.
These conductance matrices are obtained by solving standard spin diffusion equations
which are summarized in S.1A and contain all the physics of spin diffusion in one dimension noting that in non-magnetic materials there is no distinction between $x$,$y$ and $z$ components. The final ingredient (III) is $G^{ch}$, which we assume to be non-magnetic in this paper and therefore it is adequately described using same matrices as $G^{lead}$.\\
\begin{figure}[t]
 \centering
  \includegraphics[width=4cm, height=3cm]{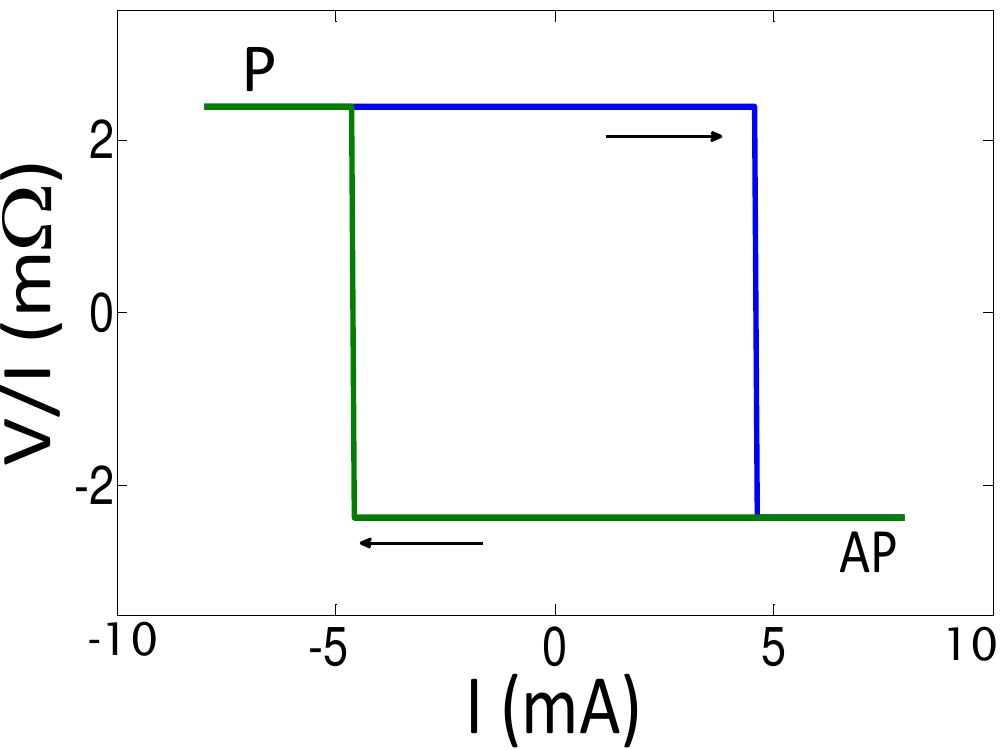}\\
\caption{Calculated spin-valve signal vs. input current closely matches the experimental results in Ref.[7]}
  \label{fig2}
\end{figure}
\indent 
The spin components $\vec{I}_{ns}$ of the current $\left[I_n\right]$ (Fig.\ref{fig1}b) into magnetic contact `$n$' provide the spin-torque that enters the LLG equation describing the dynamics of magnet `$n$':
\begin{equation}
\label{LLG}
\frac{d\hat{m}_n}{dt}=-|\gamma|\hat{m}_n\times\vec{H}
+ \alpha\hat{m}_n\times\frac{d\hat{m}_n}{dt}
- \frac{1}{qN_s}\hat{m}_n\times\left(\hat{m}_n\times\vec{I}_{ns}\right)
\end{equation}
where $q$ is the charge of electron, $\gamma$ is the gyromagnetic ratio,
$\alpha$ is the Gilbert damping parameter and $N_s\equiv M_s\Omega/\mu_B$ is the net number of Bohr magnetons comprising the nanomagnet ($M_s\equiv$ saturation magnetization and $\Omega\equiv$ volume). $\vec{H} = H_K\hat{z} - H_d\hat{y}$ represents the internal `uniaxial anisotropy' and `out-of-plane demagnetizing' effective fields acting on the magnet. The last term is the spin-torque current $\vec{I}_{st}$.\\
\begin{figure}[b]
 \centering
  \includegraphics[width=6cm, height=4cm]{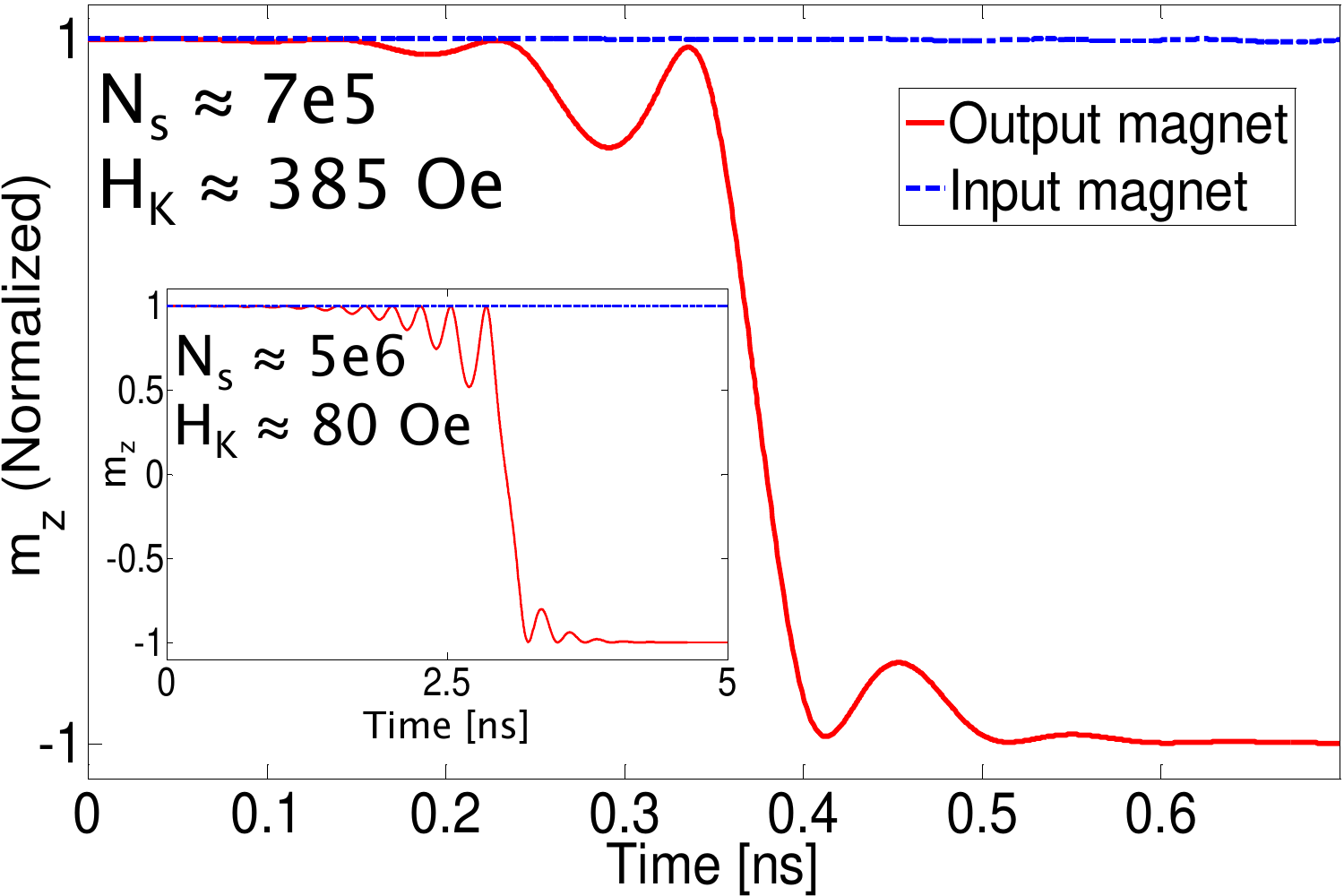}\\
  \caption{For the same current I, a magnet with smaller $N_s$ (7e5) is predicted to switch faster than one with $N_s=5e6$.}
  \label{fig3}
\end{figure}
\indent 
Figure \ref{fig3} is plotted using the self consistent model just described where the component of output magnetization along its easy axis ($z$) is shown. Here, we ignore the transit time of carriers because it is on the order of a picosecond much less than the nanomagnet dynamics (see S.2). The inset shows a magnet that has $\approx5$ times lower anisotropy field ($H_K$) and $7$ times higher $N_s$ (Here $\Omega$ and not $M_s$ has been varied to change $N_s$). It is clear that with higher $H_K$ and lower $N_s$, magnet switches faster
but this is not in the expense of higher current since $N_s$ has been reduced and it is generally accepted that switching current is proportional to $N_s$ after the work of Sun\cite{Sun}. Also, we find that with identical input and output magnets, the switching process is non-reciprocal without the need to use Bennett clocking scheme\cite{Imre,BehinAein,Bandyopadhyay} because of different voltages on the two magnets (e.g. floating output in Fig.\ref{fig1}a). This causes the output magnet to switch faster than the input magnet. We note however that the floating voltage may not be ideal for cascading circuits; such issues will be discussed elsewhere. Here we focus more on the switching energy-delay.\\
\indent 
With this in mind, it is instructive to look at how the different current components vary with time (Fig.\ref{fig4}).
It is evident that while the charge and spin currents, $I_c$ and $I_s$ continue to flow as long as a voltage is present,
the spin torque current $\vec{I}_{st}$ (c.f. Eq.\ref{LLG}) that enters the LLG equation is time-limited: it flows only during the time that the magnet is switching. Indeed we find that $\int_0^{\infty}{I^z_{st}dt} = f_1\left(2qN_s\right)$. The factor $f_1$ is exactly 1 if only a uniaxial field is present\cite{Footnote} as we might have expected from angular momentum conservation\cite{Sun,Kent} (see also S.3). However we find that $f_1$ can be less or more than 1 when fields other than uniaxial are involved (S.3). From Fig.\ref{fig4} it is evident that the total charge $Q_{tot}$ in Eqs.\ref{SW-ED} will be larger than $\int_0^{t_{sw}}{I^z_{st}dt}$ and can be written as
\begin{equation}
Q_{tot}=\int_0^{t_{sw}}Idt = \frac{I}{\overline{I_s}}f_2f_1\left(2qN_s\right)
\end{equation}
where $I$ is the charge current, $\overline{I_s}$ is the time-average spin current and $f_2=\int{I_sdt}/\int{ I_{st}^zdt}$ is a factor reflecting the fact that the spin current $|I_s|$ is somewhat larger than $|I_{st}|$ that enters Eq.\ref{LLG}.\\
\begin{figure}
 \centering
  \includegraphics[width=6cm, height=4.0cm]{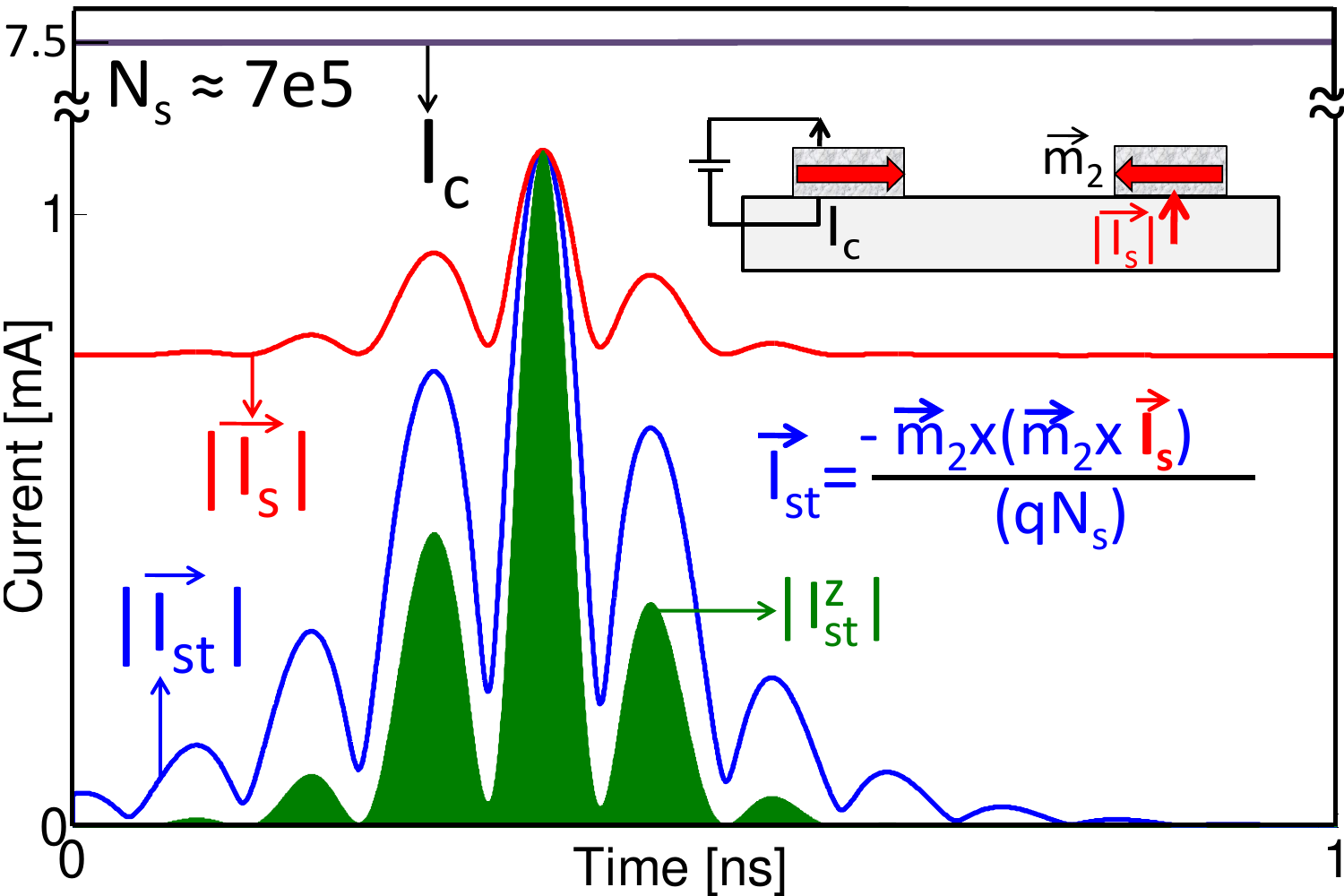}\\
  \caption{Various currents throughout a switching event.}
  \label{fig4}
\end{figure}
\indent 
Evidently, the switching energy-delay, among other things, can be improved by lowering $N_s$. While reducing $N_s$, thermal stability has to be ensured which is determined by the activation barrier ($E_b = K_u\Omega$, $K_u$ is the effective uniaxial anisotropy constant) of a magnet. $E_b$ has to be $\approx$ 10's of kT to sustain non-volatility. $N_s$ is related to $K_u\Omega$ through the following equation: $N_s = M_s\Omega/\mu_B = 2K_u\Omega/\mu_BH_K$. Taking stability into account, ultimate scaling requires magnetic materials (see e.g. Weller et al.\cite{Weller}) with high anisotropy fields ($H_K=2K_u/M_s$) where only several thousand Bohr magnetons ($N_s$) can collectively give rise to stable magnets. In short, lowering $N_s$ will keep the switching current low and high $H_K$ will decrease the switching delay. Although making devices from high anisotropy magnetic materials\cite{Kent,Weller} could have experimental challenges, we believe the underlying potential impact on lowering energy-delay of spin-torque switching remains valid. On the other hand with magnetic field switching used in schemes such as MQCA\cite{Imre}, higher $H_K$ would require a higher switching field; hence higher switching current. Similarly, the scaling of switching energy-delay based on $H_K$ and $N_s$ is not favorable to multiferroic switching of magnets\cite{Bandyopadhyay} either since higher $K_u$ and/or lower delay would require higher switching voltages. Note that presently the low voltage operation of ASL devices is offset by the large total charge ($Q_{tot}\approx2.4e7$q in Fig.\ref{fig4}), arising from a combination of large magnets and low switching efficiency. If these numbers (i.e. $N_s$ and $f_1f_2I/\overline{I_s}$) can be reduced, the advantages of low voltage operation (less parasitic capacitance and stray charge) and non-volatility (less leakage) would make ASL look attractive.\\
\indent 
In summary, we have presented a model that combines the physics of spin transport with that of nanomagnet dynamics that agrees well with available experimental data. Using this model we investigate the switching of ASL devices and show how the energy-delay scales with $N_s$ (Eqs. 1 and 4). It is also shown that for identical input/output magnets, switching can be non-reciprocal based on the applied voltages. Suitable cascading schemes will be discussed elsewhere.\\
\indent This work was supported by the institute for nanoelectronics discovery and exploration (INDEX). The authors would like to thank Y. Otani and T. Yang for sharing experimental data.

\normalsize

\section{Model}

\subsection{Spin Diffusion Equations}
Spin diffusion equations for 1-dimensional transport can be written in terms of distributed resistances $r_u$, $r_d$
and spin-flip conductances $g_{sf}$ as:
\begin{subequations}
\begin{equation}
\label{PotentialDrop}
r_uI_u = - \frac{d(\mu_u/-q)}{dx}, r_dI_d = - \frac{d(\mu_d/-q)}{dx}
\end{equation}
\begin{equation}
\label{Continuity}
\frac{dI_u}{dx} = - \frac{g_{sf}(\mu_u - \mu_d)}{-q} = - \frac{dI_d}{dx}
\end{equation}
\end{subequations}
where $\mu_u$,$\mu_d$ are the quasi-Fermi levels for up and down spins.
Defining the charge $c$ and $z$-components of voltage and current as:
$V_c\equiv\left(\mu_u+\mu_d\right)/(-2q), I_c = I_u + I_d$
and  $V_z\equiv\left(\mu_u-\mu_d\right)/(-2q), I_z = I_u - I_d$,we can rewrite the spin diffusion
equations (\ref{PotentialDrop}) and (\ref{Continuity}) as:
\begin{subequations}
\begin{equation}
\label{1DSpinDiffusion}
\frac{d^2V_c}{dx^2} = \left(r_u-r_d\right)g_{sf}V_z, \text{   }\frac{d^2V_z}{dx^2} = \left(r_u+r_d\right)g_{sf}V_z
\end{equation}
\begin{eqnarray}
\label{1DIc}
I_c &=& - \left(\frac{1}{r_u} + \frac{1}{r_d}\right)\frac{dV_c}{dx} - \left(\frac{1}{r_u} - \frac{1}{r_d}\right)\frac{dV_z}{dx} \\
\label{1DIz}
I_z &=& - \left(\frac{1}{r_u} - \frac{1}{r_d}\right)\frac{dV_c}{dx} - \left(\frac{1}{r_u} + \frac{1}{r_d}\right)\frac{dV_z}{dx}
\end{eqnarray}
\end{subequations}
The conductance matrices $G^{sh}$ and $G^{se}$ of the main paper are obtained by solving Eqs.(\ref{1DSpinDiffusion}) for a section of length $\ell$ with specified values of $\Delta V_c$ and $\Delta V_z$  and then calculating $I_c$  and  $I_z$ at either end from Eqs.(\ref{1DIc},\ref{1DIz}). The conductance matrices contain all the physics of spin diffusion in one dimension with $\lambda$ related to the distributed resistance and spin-flip conductance $g_{sf}$ by $\lambda^2 = \frac{1}{(r_u+r_d)g_{sf}}$. The advantage of using charge and spin (rather than $u$/$d$) components is that the conductance matrices are trivially extended to 4-component quantities noting that in non-magnetic materials there is no distinction between $x$,$y$ and $z$ components.\\
\subsection{Analytical Calculations}
We would like to point out that the agreement (Fig.2 of the main paper) with experiment (Fig.3d of Ref.[1] same as Ref.[7] of the main paper) could be expected from a simple steady state spin-transport calculation, as noted in the experimental papers. However, the real value of the full model described in the paper is twofold:
\begin{itemize}
\item Our  self consistent model couples spin transport and temporal dynamics of the magnet, thus enabling us to extract the switching delay. Knowledge of delay is essential to the estimation of switching energy and further, to establish the scaling with $N_s$. However, the delay cannot be obtained from steady state spin-transport calculations.
\item Without coupling the magnet dynamics to spin transport, it is impossible to demonstrate the possibility of non-reciprocal switching as was noted in the paper. The analysis of any logic switch
based on the ASL device needs to solve the spin transport model with the dynamics of input-output magents self-consistently to analyze non-reciprocity.
\end{itemize}

\emph{Steady state calculation}: Here we outline simple estimates for the steady state calculations. In Fig 2, the two measured values of the non-local resistance (output voltage per unit current) correspond to collinear (parallel and anti-parallel) configuration of the magnets for which one can obtain an analytical expression using the spin-diffusion equations presented in the previous section:
\begin{equation}
    \frac{V_{2c}}{I_{1c}} = R_{sN}\frac{e^{-\ell/\lambda_N}x_1x_2P_1P_2}{\left(1+x_1\right)\left(1+x_2\right) - e^{-2\ell/\lambda}}
\end{equation}
where
\begin{equation}
      x_1\equiv\frac{2R_1}{R_{sN}\left(1-P_1^2\right)} \hspace{0.5 cm},\hspace{0.5 cm}
      x_2\equiv\frac{2R_2}{R_{sN}\left(1-P_2^2\right)};
\end{equation}
\begin{table*}
\caption{List of material and geometric parameters for non-local spin valve structure of Ref.1}
  \centering
  \begin{tabular}{l c c c c c}
  \hline\hline
  &    Area     &    Length    &    Resistivity      &    Spin-flip length      &
  Spin resistance      \\

  &   $nm2$    &     $nm$    &    ($\rho$) $\Omega-nm$                &          ($\lambda$)$nm$                      &
                               ($R_s=\rho\lambda/A$)  $\Omega$   \\
  \hline
  Input magnet (NiFe)    &    $75\times170$        &    20    &   170     &   5     &    0.07       \\
  Input lead (Au)     &    $75\times170$        &    50   &   70      &   10    &    0.05       \\
  Output magnet (NiFe)   &    $80\times170$        &    4     &   170     &   5     &    0.06       \\
  Output lead (Au)    &    $80\times170$        &    50   &   70      &   10    &    0.05       \\
  Channel (Cu)        &    $170\times65$        &    $\ell= 270$       &   7     &   1000    &   0.63   \\
  Left lead  (Cu)     &    $170\times65$        &    $\ell_1= 10000$   &   7     &   1000    &   0.63   \\
  Right lead  (Cu)    &    $170\times65$        &    $\ell_2= 10000$   &   7     &   1000    &   0.63   \\
  \hline\hline
  \end{tabular}
  \label{tab1}
\end{table*}
$\rho_{F(N)}$ and $\lambda_{F(N)}$ are the resistivities and the spin diffusion lengths of the Ferromagnet (channel). $R_{sN} =\rho_N\lambda_N/A_N$ and $R_1, R_2 =\rho_F\lambda_F/A_F$ are the resistances over one spin-flip length for the channel and the input and output Ferromagnets respectively. P1(2) is the polarization of the input (output) magnet. Equation 1 is in agreement with the results obtained by other authors (see e.g., the analysis of lateral spin valves by Takahashi and Maekawa~\cite{Maekawa}). Using the parameters listed in Table 1, we calculate the non-local resistance to change by $\approx\ 4 m\Omega$ as observed experimentally and computed from our model.\\
\indent The critical current at which the output magnet switches can also be estimated analytically as noted in the experimental
paper. For collinear magnets we can write the spin current in the output magnet as
\begin{equation}
    - \frac{I_{2s}}{I_{1c}} = \frac{e^{-\ell/\lambda}x_1P_1}{\left(1+x_1\right)\left(1+x_2\right) - e^{-2\ell/\lambda}}
\end{equation}
which is approximately 10$\%$ for the parameters in the experiment. For the output magnet, using the expressions obtained by Sun~\cite{Sun}, we estimate a critical spin current of
\begin{equation}
    I_{s,cr} = \frac{2q}{\hbar}\alpha\left(2Ku\Omega + K_d\Omega)\right) = 0.46mA
\end{equation}
Here we used a Gilbert damping parameter $\alpha=0.007$ and estimated $K_u=H_KM_s/2$ and $K_d=2\pi M_s^2$ using
$H_K = 80$ Oe, $M_s = 780$ $emu/cc$. Noting that the input current is approximately 10 times
the spin current delivered to the output magnet, a switching current of 4.6 mA is comparable to that observed in the experiment as well as Fig.1 of our paper. Also note that the experiment is performed at $10$K so the zero temperature estimations from LLG are relatively accurate.\\

\subsection{Parameters used for the experimental benchmark of Fig.2.}
As mentioned earlier in the main paper, our model describes real structures fairly well as it mimics the switching characteristics of the experiment of Yang et.al.\cite{Otani2008} (Fig. 2). The material and geometric parameters used to benchmark our model are summarized in Table 1. These parameters have been obtained from experimental papers~\cite{Otani2008,Yang2008} and private communications with the authors. Apart from the parameters listed in the table, we have to point out that one also needs the value of the so called
mixing conductances\cite{Brataas} which give the $x-y$ block of the interface conductance matrix $G^{int}$ mentioned in our paper. These
values generally are calculated using ab inito models. We use quantities provided by Brataas et al.\cite{Brataas} noting that they can be estimated using $g_{mix}=g_Qk_f^2A/4\pi$~\cite{Bratas02} where $g_Q$ is the quantum of conductance, $k_f$ is the wave-vector in the channel and A is the area of the magnet-channel interface. Note that $G^{int}_{xy}$ and $G^{int}_{yx}$ are the matrix elements that give rise to the field-like term and it is generally believed that in such metallic structures they are negligible. For a precise match, the polarization of the NiFe/Cu magnet channel interface has been assumed to be 0.5, well within the experimentally measured range\cite{Yang2008}.
\section{Transit Time}
Here we would like to estimate the transit time of carriers due to the diffusion in channel
from input magent to the output magnet.
As we will see shortly, this time scale is much faster than nanomagnet dynamics and because of this,
the time of flight of carriers has been neglected in our models presented in the paper.
This time of flight can be estimated as:
\begin{equation}
t_{\text{transit}} = \frac{L^2}{2D}
\end{equation}
Where $L$ is the distance traveled and $D$ is the diffusion coefficient.
We set $D = v_F\lambda_{\text{mfp}}$. $v_F=1.57\times10^6 \text{m/s}$ is the Fermi velocity for copper\cite{AshcroftMermin}
and $\lambda_{mfp}$ is the mean free path which can be calculated using:
\begin{equation}
\lambda_{\text{mfp}} = \frac{\sigma mv_F}{nq^2}
\end{equation}
where $\sigma=5.8\times10^7\Omega^{-1}m^{-1}$ is the copper conductivity,
$m=9.11\times10^{-31}$kg is the free electron mass, $n = 8.47\times10^{28}/m^3$ and $q=1.6\times10^{-19}$C is the electron charge\cite{AshcroftMermin}. Using these numbers we get $\lambda_{\text{mfp}}\approx38$ nm. One could also use
a more conservative estimate of $\lambda_{\text{mfp}}\approx10$ nm for less pure copper. We can now calculate the time it takes for the carriers to traverse a distance of $L=100$ nm in copper:
\begin{eqnarray}
\nonumber
&t_{transit} = \frac{L^2}{2\lambda_{\text{mfp}}v_F} = \frac{\left(100\times10^{-9}\right)^2}{2\lambda_{\text{mfp}}\left(1.5\times10^6\right)}&\\
\nonumber
&=\left\{
\begin{array}{rcl}
1.7\times10^{-13}sec & \text{for} & \lambda_{\text{mfp}}=38 nm\\
6.4\times10^{-13}sec & \text{for} & \lambda_{\text{mfp}}=10 nm
\end{array}
\right.&
\end{eqnarray}
These quantities clearly show that the dominant characteristic times are not the transit time of carriers, but rather they are that of nanomagnet dynamics, therefore at this time they have not been included in our models.
\section{Conservation of angular momentum}
\subsection{Case 1: No out-of-plane demagnetizing field. ($H_d=0$)}
We are interested in the time-integral of LLG equation throughout switching:
\begin{eqnarray}
\nonumber
\label{}
\int_0^{t_{sw}}{\frac{d\hat{m}}{dt} dt} &=& -|\gamma|\int_0^{t_{sw}}{\hat{m}\times\vec{H} dt}
+ \int_0^{t_{sw}}{\alpha\hat{m}\times\frac{d\hat{m}}{dt} dt}\\
\nonumber
&-& \int_0^{t_{sw}}{\frac{1}{qN_s}\hat{m}\times\left(\hat{m}\times\vec{I}_{ns}\right) dt}
\end{eqnarray}
For the component of magnetization along its easy axis ($z$) we have:
\begin{eqnarray}
\nonumber
\label{LLGzInt}
\int_0^{t_{sw}}{\frac{dm_z}{dt} dt} =& -|\gamma|\int_0^{t_{sw}}{ \left[\hat{m}\times\vec{H}\right]_z dt}&\\
\nonumber
&+ \int_0^{t_{sw}}{\alpha\left[\hat{m}\times\frac{d\hat{m}}{dt}\right]_z dt}&
- \int_0^{t_{sw}}{\frac{I^z_{st}}{qN_s} dt}
\end{eqnarray}
In the absence of any out-of-plane demagnetizing field, the precessional
term ($\hat{m}\times\vec{H}$) has no $z$-component; hence the first term
on the right hand side vanishes. For the second term we have:
\begin{equation}
\nonumber
\label{}
\int_0^{t_{sw}}{\alpha\left[\hat{m}\times\frac{d\hat{m}}{dt}\right]_z dt} = \oint_0^{t_{sw}}{m_xdm_y} - \oint_0^{t_{sw}}{m_ydm_x}\equiv 0
\end{equation}
Each integral on the right hand side of above equation amounts to zero because
it is an integration of a total differential over a closed path. Noting the left hand side of
Eq.\ref{LLGzInt} for a switching event ($\pm z$ to $\mp z$) amounts to $\pm2$ we have:
\begin{equation}
\nonumber
\int_0^{t_{sw}}{\left|I^z_{st}\right| dt} = q2N_s
\end{equation}
\subsection{Case 2: Non-zero Out-of-plane demagnetizing field. ($H_d= - 4\pi M_s m_y\hat{y}$)}
For the case where both the in-plane uniaxial field and the out-of-plane demagnetizing
field are also present the first term on the r.h.s of Eq.\ref{LLGzInt} has a non-vanishing component:
\begin{equation}
\label{}
\nonumber
\int_0^{t_{sw}}{ \left[\hat{m}\times\vec{H}\right]_z dt} = -4 \pi M_s\int_0^{t_{sw}}{m_xm_y}dt
\end{equation}
Note that this term can amount to net positive or negative values. This depends
on various conditions such as the ratio of the uniaxial anisotropy field to the demagnetizing
field and the amount of current overdrive. Based on the simulations that we have performed,
we see the following range:
\begin{equation}\label{CONSERV2}
\nonumber
\int_0^{t_{sw}}\left|I^z_{st}\right|dt = f_1q\left(2N_s\right), \hspace{0.5cm} f = 0.5 - 1.8
\end{equation}
which never the less gives a value on the order of $2N_s$ for a switching event.

\end{thebibliography}


\begin{thebibliography}{100}
\bibitem{Footnote} Also a voltage pulse of length $t_\text{pulse}$ could ensure that the magnet switches, even though $\int_0^{t_\text{pulse}}I^z_{st}<f_1\left(2qN_s\right)$. But we are ignoring this aspect.
\end{thebibliography}
\end{document}